\documentclass[global,twocolumn]{svjour}
\usepackage{latexsym}
\usepackage{graphics}
\journalname{Applied Physics B}
\begin{document}
\title{Very long optical path-length from a compact multi-pass cell}
\author{Dipankar Das\thanks{dipu@physics.otago.ac.nz}
and Andrew C. Wilson
}

%
\institute{Jack Dodd Centre, Department of Physics, University of Otago, PO Box 56, Dunedin 9016, New Zealand}
\date{Received: date / Revised version: date}
%
\maketitle
\begin{abstract}
The multiple-pass optical cell is an important tool for laser absorption spectroscopy and its many applications.  For
most practical applications, such as trace-gas detection, a compact and robust design is essential.  Here we report
an investigation into a multi-pass cell design based on a pair of cylindrical mirrors, with a particular focus on
achieving very long optical paths.  We demonstrate a path-length of 50.31 m in a cell
with 40 mm diameter mirrors spaced 88.9 mm apart - a 3-fold increase over the previously reported longest path-length
obtained with this type of cell configuration. We characterize the mechanical stability of the cell and describe the
practical conditions necessary to achieve very long path-lengths.
\end{abstract}

\section{Introduction}
Multiple-pass optical cells are ubiquitous in laser spectroscopy laboratories,
providing the long optical path-lengths necessary to study weak trace-gas
absorption features in spectral regions
where there are excellent laser sources available. Investigations where multi-pass cells have been used cover
a very broad range of topics, including environmental monitoring \cite{WMS93,LFW2000,SIW02},
combustion processes \cite{SAN80,MCD01}, medical diagnostics \cite{GIS10},
and fundamental atomic and molecular physics \cite{ZLW07,QUG08}. Multi-pass cells
are used when the absorption length required for single pass detection is inconveniently
long, but path-lengths on the kilometer scale provided by high-finesse optical cavities are not required.
In general, multi-pass cells are relatively compact and simpler
to work with than the high-finesse optical
cavities used in cavity ring-down spectroscopy \cite{AFM84,WNO98}. For example,
the latter usually requires spatial mode matching, precise optical alignment
and resonant excitation.

The design of a compact multi-pass cell is of considerable interest
in the development of portable gas sensors, such as for monitoring
and detecting the presence of noxious gases. Many important
trace-gases have very weak absorption features at wavelengths
where low-cost laser sources are now readily available.
Compact, long-path, multi-pass cells therefore have the
potential to greatly increase the range of species for which
laser-based detection is practical. The motivation for this is
that currently industrial gas sensing is mostly done with electro-chemical
sensors for which stable calibration is a major problem.

The most commonly used multi-pass cells are the White cell \cite{WHI42,WHI76,CHB91}, or
the Herriott cell with either spherical mirrors or custom-made astigmatic mirrors
\cite{HKK64,HSC65,HSC265,MKZ95,MKK89,RKW02}. Cells based on spherical
mirrors are straightforward to construct and produce a simple beam pattern.
However, these cells are not especially compact, because achieving a long
and re-entrant path requires large mirrors. In addition, because the light
pattern on the spherical mirrors consists of a circular ring of spots, the light
fills only a small fraction of the cell's volume. In this sense these cells
are not space efficient. More dense patterns can be produced with astigmatic
mirrors, but this requires customized mirrors with extraordinarily well defined focal lengths.

In 2002, Hao \emph{et al.} \cite{HQW02} developed a model for an optical cell based upon two standard cylindrical
mirrors and presented an experimental implementation of this scheme.  In their model they set the mirror axes
orthogonal to one another and predicted Lissajous type-patterns.  Their
experimental results validated the predictions of the model, but they also included an image of a more complex
spot pattern obtained by rotating the mirrors with respect to one another.  More recently, Silver
\cite{SIL05,SILP05} investigated the behaviour of a multip-pass cell based on cylindrical mirrors as one of the
mirrors is rotated with respect to the other, revealing a
range of dense and complex beam patterns. Silver presented a
model describing the system, and a map of the allowed re-entrant
pass number as a function cell geometry (spacing and twist angle).
The map shows that many very-long-path configurations are
possible, but also that re-entrant solutions are often very closely
located in parameter space. From a practical perspective, a key question for this
scheme is, what is the maximum stable path-length that one can reasonably
achieve? If
mechanical instability causes the cell to drift between
neighboring re-entrant solutions, the path-length will change,
making absolute spectroscopic measurements impossible. In this
paper we investigate this issue.

To date, there have been only two laboratory measurements reported in the
literature that use the Silver scheme. First, Silver himself reported an O$_2$ measurement using a pair
of 5 cm square cylindrical mirrors that gave a 6.35 m intra-cell path-length
with a mirror spacing of 4.47 cm (142 passes). Second, Kasyutich \emph{et al.} \cite{KAM07}
reported a path-length of 15.18 m using 76 mm diameter cylindrical mirrors spaced by
110 mm (138 passes). In this paper we report for the first time, path-lengths
up to 50 m - a 3-fold increase over the longest path-length previously reported.
Our cell is very compact, using cylindrical mirrors with a diameter of only
40 mm, and spaced approximately 90 mm apart (556 passes). To confirm the useful
operation of our cell configuration we perform a simple wavelength modulation spectroscopy
measurement on carbon monoxide at 1565 nm. \label{intro}

\begin{figure}
\resizebox{0.4\textwidth}{!}{\includegraphics{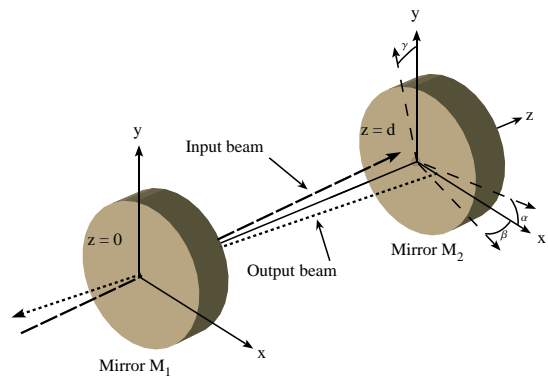}}
\caption{(Color on line) Geometry of the cell with two cylindrical
mirrors.}
\label{geometry}
\end{figure}

The basic geometry of the multi-pass cell configuration is shown in Fig.
\ref{geometry}, where \emph{$\alpha$} is the angle of rotation of the
mirrors about the \emph{z} axis (the twist angle), and \emph{$\beta$} and \emph{$\gamma$} are
respectively the horizontal and vertical tilt angle measured with respect
to the \emph{x} and \emph{y} axis. The front cylindrical mirror M$_1$ has a
central entrance/exit hole and the principal radius
of curvature is aligned with the \emph{y} coordinate axis. Initially, the
rear cylindrical mirror M$_2$ is aligned with the \emph{x} coordinate
axis (a rotation angle $\alpha = 0$). The centers of the mirrors
M$_1$ and M$_2$ are located at $\emph{z = }0$ and $\emph{z =} d$ respectively.
A laser beam is injected into the cell at \emph{x = y = z =} 0 at an
angle $\sim$8$^\circ$ from the cell axis in the \emph{xz} plane as shown in Fig. \ref{geometry}.
 Under optimal operating conditions, the path will be re-entrant so that after \emph{N}-1 reflections on
the mirrors (or \emph{N} passes), the beam will propagate out of the cell at an
angle $\sim$8$^\circ$ in the \emph{xz} plane and at $\emph{x = y = z =} 0$ \cite{KAM07}.
In order to tune to a very long optical path-length, we rotate mirror M$_2$ to a predetermined angle $\alpha \neq 0$. As described in earlier
works \cite{SIL05,KAM07}, for a given radius of curvature
and distance \emph{d} between the mirrors, $\alpha$ can be calculated using
the ray matrix formalism \cite{KOL66}. \label{sec:2}

\section{Experimental}
\begin{figure}
\vspace{1cm}
\resizebox{0.4\textwidth}{!}{\includegraphics{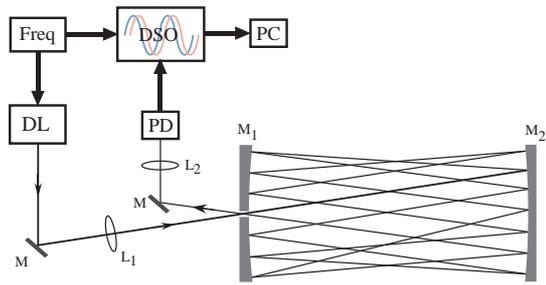}}
\caption{Experimental optical setup: DL - Diode laser, PD - Photodiode,
Freq - Frequency generator, DSO - Digital Oscilloscope, PC - Computer,
M - Mirrors, M$_{1,2}$ - Cylindrical mirror, L$_{1,2}$ - Lens.}
\label{expsetup}
\end{figure}

\begin{figure}
\resizebox{0.35\textwidth}{!}{\includegraphics{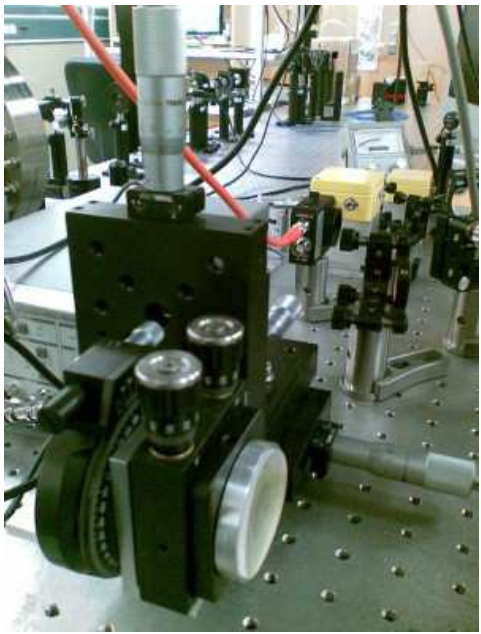}}
\caption{(Color on line) One of two mirror mount assemblies.}
\label{foto}
\end{figure}

Our optical setup is shown in Fig. \ref{expsetup}. The cylindrical mirrors used in the multi-pass
cell have a 200 mm radius of curvature, a 40 mm diameter, and a thickness of 5.2 mm. The front cylindrical
mirror M$_1$ has a 2 mm diameter central hole through which light is injected and returned from the cell.
Both M$_1$ \& M$_2$ have a broadband dielectric coating with a high reflectivity (99.64\% average value). Since
the light intensity at the output reduces exponentially with the number of reflections, high-reflectivity mirrors
are necessary to achieve very long path-lengths with sufficient light exiting the cell for spectroscopic measurements.
The mirrors are mounted on stages which provide precise control over the pitch \emph{$\beta$}, yaw \emph{$\gamma$},
twist \emph{$\alpha$}, and translation (Fig. \ref{geometry}). Gimbal mirror mounts are attached to micrometer-adjustable rotation stages for
fine rotation control. In turn, the precision rotation stages are fixed to \emph{XYZ}
translation stages for fine adjustment of displacement. The assembly of three mounts is shown in Fig. \ref{foto}. As we will show,
such sophisticated (and bulky) mounting is not necessary in order to achieve very long path-lengths, but it allows us to investigate tolerance
of the scheme to mechanical misalignment.

The laser source is a collimated 1565-nm VCSEL (vertical-cavity surface-emitting laser) diode
laser (Vertilas VL-1570-1-SP-A5, 2 mw @ 7.5 mA), which has an internal TEC (thermo-electric cooler)
unit for temperature control.  The lens L$_1$ (focal length 200 mm) in the input beam path is used to produce a 0.14 mm beam waist at the
cell's 2 mm diameter entrance/exit aperture. For re-entrant solutions, as the light propagates within
the cell its diameter varies, but overall expansion is compensated by mirror focusing. When we place
a beam waist at the cell entrance, we observe an identical waist at the aperture as the beam exits the cell.
 This feature helps considerably to avoid the problem of the beam clipping the aperture as it enters/exits the cell,
so that the many re-entrant solutions with similar (and tightly spaced) spatial patterns on the front
mirror can be spatially resolved. The lens L$_2$
(focal length 100 mm) in the output beam path focuses the light onto a photodiode.

\begin{figure}
\vspace{1cm}
\resizebox{0.4\textwidth}{!}{\includegraphics{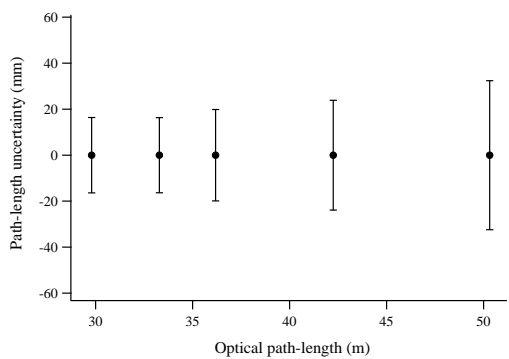}}
\caption{Uncertainties in the path-length as measured using the phase-shift method.} \label{error}
\end{figure}

The 1565 nm near-infrared (NIR) light we use is invisible, and our dielectric mirror coatings
have low reflectivity at visible wavelengths, so we have developed a new strategy for achieving
very long path-lengths that does not involve observing the beam spot pattern on the mirrors.
This scheme relies on the fact that we can align the cell's spacing, as well as mirror
displacements and tilts, for a small number of passes, and then simply rotate one of
the mirrors to ``dial up'' longer path-lengths. For longer path-lengths, small adjustments
are necessary, but straightforward. When properly aligned we have up to 556 passes, so
aside from the problem of imaging NIR light, we cannot simply count the number of spots
on the mirrors to determine the path-length. To solve this problem we use intensity modulation
and measure the phase shift induced by cell's optical path. The intensity modulation is
induced by applying a small amplitude modulation to the diode laser's injection current using
a bias tee (Mini-Circuits ZFBT-4R2GW+). The phase difference $\phi_{L}$ between the input
and output beams is given by $\phi_{L}=\frac{2\pi\nu}{c}L$, where \emph{L}, \emph{$\nu$}
and \emph{c} are respectively the traversed path, modulation frequency and velocity
of light. The output beam from the multi-pass cell is focused by a lens onto a high
bandwidth photodiode (Thorlabs PDA10CF-EC) and the phase is measured with an oscilloscope.
The measured phase shift includes contributions from the optical path external
to the cell and also from electrical cables. To measure this phase offset, we bypass the cell by placing a mirror
in front of the cell which directs the beam onto the photodiode. The optical path-length
is then $L=\frac{c}{2\pi\nu}\{\phi_{measured}-\phi_{offset}\}$. In this method the optical
path-length is only known to within a distance associated an integer multiple of 2$\pi$
phase shift. Therefore to ensure that our path-length measurement is correct, we step the modulation
frequency from 1 to 80 MHz and measure the phase shift change. The correct path-length is the
one that is consistent across all frequency steps, and 80 MHz provides sufficient
precision to determine the path-length within an integer number of passes.
Using the oscilloscope to average 32 phase measurements at each modulation frequency,
we obtain an uncertainty of $\pm$0.03 m in the 50 m path. The absolute uncertainty in path-length increases slightly
for longer paths, as shown in Fig. \ref{error}, but the relative uncertainty remains almost constant. In all cases our distance measurement uncertainty
is considerably less than the mirror spacing, so that our measurements give us precisely
the number of passes within the cell.
\label{sec:3}

\section{Results and Discussion}
\begin{figure}
\resizebox{0.4\textwidth}{!}{\includegraphics{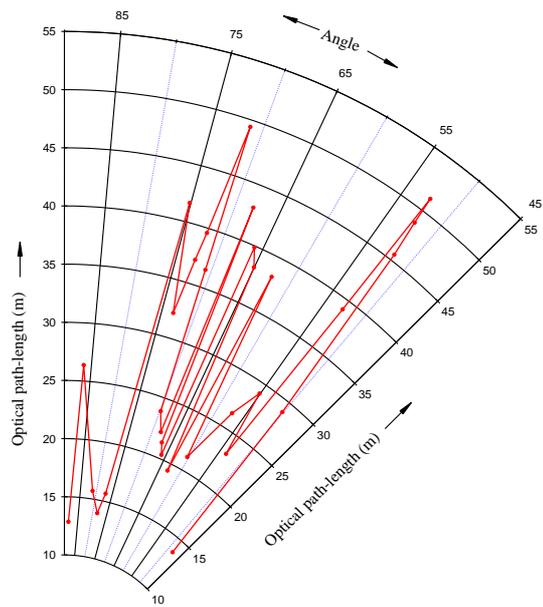}}
\caption{(Color on line) Dependency of optical path-length on the twist angle
for a cell base length of $\sim$ 90 mm.}
\label{polar}
\end{figure}

We begin optical alignment by setting the spacing between the cylindrical mirrors (the cell base length)
to the focal length of each mirror ($\sim$100 mm) and rotate the mirrors so that their axes are orthogonal.
At an $\sim$8$^\circ$ angle of incidence (and with very
little effort) this preliminary alignment of the cell generates a 5 m optical
path-length. We then carefully rotate M$_1$ to change the twist
angle between M$_1$ and M$_2$, and seek re-entrant solutions. The result of this procedure
is shown in Fig. \ref{polar}. In order to focus on long paths, in this figure we only show
results for path-lengths greater 10 m. For a complete 90$^\circ$ rotation of M$_1$, we obtain
many re-entrant solutions with optical path-lengths ranging from 3 to 50 m. As can be seen
from Fig. \ref{polar}, a path-length slightly less, and slightly more, than 50 m can be
obtained for a twist angle of $\sim$53$^\circ$ and $\sim$71$^\circ$ respectively.
Furthermore, optical path-lengths in the range 35 to 45 m appear frequently for twist angles
between $50^\circ$ to $75^\circ$.

For very long path-lengths (i.e. those approaching 50 m), it is not surprising that the
re-entrant solutions become relatively sensitive to aspects of the alignment. Table \ref{tab:1}
shows the range over which the cell alignment parameters can be changed while still maintaining
the 50 m re-entrant path.  The tolerance values correspond to changes
that result in a 50\% reduction in the light intensity exiting the cell. From Table \ref{tab:1}
we conclude that the scheme is relatively insensitive to changes in transverse displacement of the mirrors
($\Delta$\emph{x} and $\Delta$\emph{y}), and to changes in pitch and yaw ($\Delta\gamma$ and $\Delta\beta$).
The sensitivity to mirror tilt is due to the beam pattern walking off the mirror coatings (which,
if necessary, could be avoided with larger diameter mirrors). The scheme is, however, highly sensitive
to changes in the mirror spacing $\Delta$\emph{z} and twist angle $\Delta\alpha$). If the mirror spacing
changes by only ±0.02 mm, or the twist angle changes by ±0.02$^\circ$, the output beam becomes clipped by the
2 mm diameter entrance/exit aperture.

To put the mirror spacing requirement into perspective, a 0.02 mm change in a 90 mm length cell
constructed with aluminium or stainless steel corresponds to a temperature
change of only approximately 10 K. This is fine for typical laboratory conditions, but is unlikely to be
acceptable for industrial conditions. In harsh thermal environments our cell would have to be constructed
of low thermal expansion material, such as Invar, or require active temperature stabilization. A more practical
solution that we have implemented is to use polymer materials which have extremely low (a few ppm) thermal
expansion along one direction. Thermal expansion perpendicular to this direction is considerably larger,
but this is unimportant for our purpose.  As for the sensitivity to twist angle, if the mirrors were to be
spaced by a solid material, linear thermal expansion will not lead to a twist. However, if the spacer material
were to have significant internal stress, this might cause a tilt problem, and perhaps this issue deserves
further investigation.

\begin{figure}
\vspace{1cm}
\resizebox{0.4\textwidth}{!}{\includegraphics{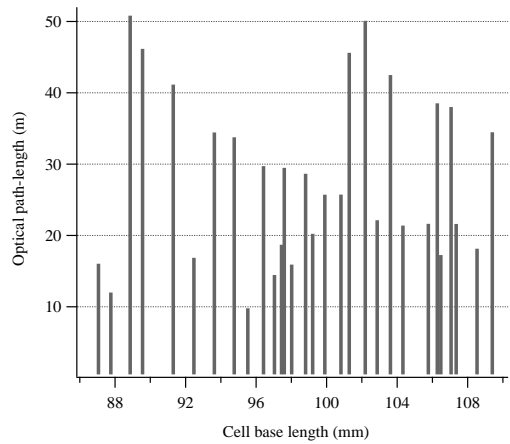}}
\caption{Optical path-length dependency on the cell base length
for a twist angle $\sim$71$^\circ$.}
\label{cellbaselength}
\end{figure}

To investigate the dependency on mirror spacing, we measured the optical path-length of successive re-entrant
solutions as we varied the base length from 85 mm to 100 mm, keeping the mirror twist angle at 71$^\circ$.
Results are given in Fig. \ref{cellbaselength}. We observe path-lengths in excess of 50 m for cell base
lengths of 88.9 mm and 102.2 mm. As expected, we observe very many re-entrant solutions with long
path-lengths. Note that, as with the previous figure, we have not included
re-entrant paths with lengths less than 10 m.

\begin{figure}
\vspace{1cm}
\resizebox{0.4\textwidth}{!}{\includegraphics{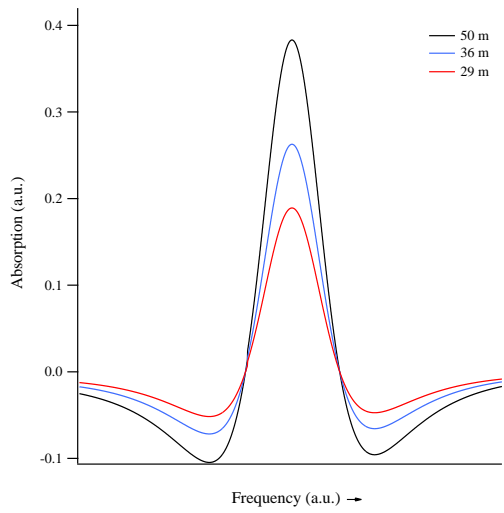}}
\caption{(Color on line) 2$f$ WMS absorption spectrum of CO near 1565 nm for different path-lengths.} \label{CO-Spec}
\end{figure}

Finally, we performed a simple 2$f$ wavelength modulation spectroscopy (WMS) measurement
on molecular CO (R-branch, 3$\upsilon$ band) at three different path-lengths. For this measurement
the entire optical setup was enclosed is a sealed box. CO was injected into the box, and the concentration was measured
with an electrochemical sensor to be approximately 40 ppm, which is a factor of 4 above the standard
safe-exposure limit. 2$f$ WMS absorption spectra for the different path-lengths are shown in
Fig. \ref{CO-Spec}. The spectra were recorded for 29.79 m, 36.18 m and 50.31 m path-lengths respectively.
We started with the 29.79 m path-length and the higher path-length spectra were obtained with subsequent
rotation of the M$_1$. As can be seen in Fig. \ref{CO-Spec}, we obtain a good signal to noise ratio and
the absorption increases by the appropriate ratio for the weak absorption limit.
These features suggest that this multipass cell is well-suited to measurement of absolute concentration.

Two factors which can limit the performance of long path (and therefore many pass)
cells are attenuation of the light to the point where the absorption signal is too
weak, and also interference effects that lead to noise on the absorption signal.
The input light power was 1.02 mW and for the 50 m path the output power incident onto
the photodiode was 0.136 mW, giving us an average reflectivity for the cell mirrors of 99.64\%.
This was clearly sufficient to obtain good signal to noise. Also, at this signal
level we do not observe etalon effects, but at lower CO concentrations we expect that
this will ultimately limit detection sensitivity as it does in most WMS setups.
\label{sec:4}

\section{Conclusion}
Using a multi-pass cell based upon two cylindrical mirrors, very long optical
path-lengths (up to 50 m) have been demonstrated. The cell is compact, having
a volume of only 113 cm$^{3}$. A method for determining the optical path-length
based on intensity modulation was described. This method avoids the need to image
the beam pattern within the cell. Mechanical tolerances were investigated for the
case of a 50 m path-length, and it was shown that high tolerance on the mirror
spacing and twist angle are required for robust operation. A wavelength modulation spectroscopy measurement
was performed on a weak carbon monoxide transition in the near infrared. Even with 556
passes, typical dielectric mirror coatings gave sufficient light output to obtain good WMS signal to noise.

We thank and Alex Barker, Andrew Wade, Hamish Findlay and Cal McClean all at \emph{Photonic Innovations Ltd} for helpful
discussions, and Tom Holmes for his assistance with construction of the setup. This work
was supported by the New Zealand Foundation for Research, Science \& Technology, contract
NERF-UOOX0703.

\begin{table}
\caption{Mirror position and alignment tolerances for a $\sim$ 50 meter
path-length}
\label{tab:1}
\begin{tabular}{ccccccc}
\hline\noalign{\smallskip}
Parameter &&& M$_1$ &&& M$_2$  \\
\noalign{\smallskip}\hline\noalign{\smallskip}
$\Delta$x &&& $\pm$0.35 mm &&& $\pm$1.20 mm \\
$\Delta$y &&& $\pm$0.35 mm &&& $\pm$1.10 mm \\
$\Delta$z &&& $\pm$0.02 mm &&& $\pm$0.02 mm \\
$\Delta\gamma$ &&& $\pm$0.10$^\circ$  &&& $\pm$0.14$^\circ$  \\
$\Delta\beta$ &&& $\pm$0.10$^\circ$   &&& $\pm$0.14$^\circ$  \\
$\Delta\alpha$ &&& $\pm$0.02$^\circ$  &&& $\pm$0.02$^\circ$  \\
\noalign{\smallskip}\hline
\end{tabular}
\vspace*{1cm}
\end{table}

\end{document}